\documentclass[aps,prl,reprint,superscriptaddress,amsmath,amssymb]{revtex4-2}
\usepackage{bm,mathrsfs,hyperref}
\begin{document}
\title{Transport Determines Spectral Distance in Mesoscopic Thermodynamics}
\author{Giorgio Sonnino}
\affiliation{Universit\'e libre de Bruxelles (ULB), 1050 Brussels, Belgium}
\affiliation{International Solvay Institutes for Physics and Chemistry, 1050 Brussels, Belgium}
\begin{abstract}
We derive a distance between thermodynamic states from a positive Hermitian transport tensor within a mesoscopic canonical model. An exact first-order factorization of the evolution generator gives a commutator norm and the metric $G=2g$, where $g$ denotes the symmetric transport tensor. Gradients of the antisymmetric tensor can change the evolution, but they leave this distance unchanged. The construction allows coordinate-dependent coefficients. For constant Onsager coefficients, a Clifford representation gives the same metric. In a two-channel thermoelectric system, the ratio of normalized principal distances is $\sqrt{1+ZT}+\sqrt{ZT}$. Transport coupling thus fixes the anisotropy of this metric. The construction requires neither an independent metric nor a physical spin variable.
\end{abstract}
\maketitle

Transport rates depend on direction in thermodynamic state space. Equal coordinate displacements can drive very different fluxes. Near equilibrium, Onsager's matrix describes this dependence \cite{Onsager1931a,Onsager1931b}. We ask whether the same coefficients also determine distances between states. We use the canonical mesoscopic model of Ref.~\cite{Sonnino2024a}, with the fluctuation scale studied in Refs.~\cite{Sonnino2024b,Sonnino2025}. We assume its coordinate--force commutation relations and represent the fluxes by self-adjoint operators. A quadratic generator reproduces these fluxes through coordinate commutators. Factoring that generator into first-order operators allows us to evaluate Connes' distance formula \cite{Connes1994,GraciaBondia2001}. Only the symmetric transport tensor enters the distance. Gradients of the antisymmetric tensor can still change the evolution. We prove this separation for coordinate-dependent coefficients and recover the constant Onsager limit with a Clifford representation. We then calculate how thermoelectric coupling sets the anisotropy of the normalized two-channel metric. The distance compares points in thermodynamic-coordinate space. A dissipation cost for a finite-time protocol requires additional information, as does the affine connection of nonlinear Thermodynamic Field Theory \cite{Sonnino2009,Sonnino2016}. 

\noindent The present Letter establishes how transport determines spectral distance in mesoscopic thermodynamics. A separate manuscript \cite{Sonnino2026} investigates the associated spinorial structure, constructing an exact graded first-order factorization of the evolution generator for a positive Hermitian transport tensor.

\textit{Flux observables and evolution.} Let $\alpha^\mu$, $\mu=1,\ldots,n$, denote thermodynamic coordinates in a fixed coordinate representation. Their conjugate forces obey
\begin{equation}
 [\alpha^\mu,\widehat X^\nu]=i\varepsilon\delta^{\mu\nu},
 \qquad \widehat X^\nu=-i\varepsilon\partial^\nu
 \label{eq:canonical}
\end{equation}
where $\partial^\mu=\partial/\partial\alpha^\mu$ and $\varepsilon=\slash\!\!\! k_B/2$ is the mesoscopic fluctuation parameter of the cited framework. Repeated indices are summed. Consider
\begin{equation}
 \tau_{\mu\nu}=g_{\mu\nu}+if_{\mu\nu},\qquad
 g_{\mu\nu}=g_{\nu\mu},\quad f_{\mu\nu}=-f_{\nu\mu}
 \label{eq:transport}
\end{equation}
with real $g$ and $f$. We take the coefficients to be smooth and bounded, and assume uniform positivity of $\tau$ on $\mathbb R^n$. Under these conditions, the quadratic form below defines a nonnegative self-adjoint operator after closure. On other domains, the construction also requires compatible boundary conditions. We use $\tau$ to specify a Hermitian differential operator in the mesoscopic model. Its complex entries do not represent a classical law mapping real forces to complex measured currents. Set $A_\mu=\tau_{\mu\nu}\widehat X^\nu$ and choose $\widehat J_\mu=(A_\mu+A_\mu^\dagger)/2$. Integration by parts gives
\begin{equation}
 \widehat J_\mu=-i\varepsilon\left(g_{\mu\nu}\partial^\nu
 +\tfrac12\partial^\nu g_{\mu\nu}\right)
 -\tfrac{\varepsilon}{2}\partial^\nu f_{\mu\nu}
 \label{eq:flux}
\end{equation}
We choose this Hermitian prescription as the quantization rule; self-adjointness does not select it uniquely. Varying antisymmetric coefficients contribute to the flux, whereas constant $f$ contributes nothing. We require the coordinate commutators of a self-adjoint generator $D$ to reproduce these fluxes:
\begin{equation}
 i\varepsilon[D,\alpha^\mu]=\widehat J_\mu
 \label{eq:compatibility}
\end{equation}
The minimal divergence-form realization is
\begin{equation}
 D=-\tfrac12\partial^\mu\left(\tau_{\nu\mu}\partial^\nu\right)\quad{\rm with}\quad \tau_{\nu\mu}=\tau_{\mu\nu}^*
 \label{eq:generator}
\end{equation}
Substitution verifies Eq.~\eqref{eq:compatibility}. The commutator leaves a real multiplication potential undetermined because such a potential commutes with every coordinate. We choose the minimal realization by setting it to zero. For smooth compactly supported $\psi$,
\begin{equation}
 \langle\psi,D\psi\rangle=
 \tfrac12\int d^n\alpha\,
 (\partial^\mu\psi)^*\tau_{\mu\nu}\partial^\nu\psi\geq 0
 \label{eq:form}
\end{equation}
The closure of this form defines the self-adjoint generator used here. For the dynamics, we assume a Schr\"odinger-type mesoscopic evolution law. Requiring Heisenberg evolution to satisfy $d\alpha^\mu/dt=J_\mu$ fixes the effective generator as $\widehat\sigma_{\mathrm{eff}}=\varepsilon^2D$. The state then obeys $i\partial_t\psi=\varepsilon D\psi$. The earlier framework motivates this law through a formal time-entropy-production analogy. Here time is a parameter. We use no time operator: Eq.~\eqref{eq:compatibility} ensures that the assumed evolution gives the chosen flux observables as coordinate velocities. Unitary evolution preserves the von Neumann entropy of the full state. A thermodynamic entropy balance requires a separate entropy-production observable and a statistical description of coarse-graining. Antisymmetry removes $f$ from the principal second-order symbol because $f_{\mu\nu}\partial^\mu\partial^\nu=0$. Expanding Eq.~\eqref{eq:generator} yields
\begin{equation}
 D=-\tfrac12g_{\mu\nu}\partial^\mu\partial^\nu
 -\tfrac12(\partial^\mu g_{\mu\nu})\partial^\nu
 +\tfrac i2(\partial^\mu f_{\mu\nu})\partial^\nu
 \label{eq:expanded}
\end{equation}
Varying $f$ therefore changes the first-order part of $D$. Since $\partial^\nu\partial^\mu f_{\mu\nu}=0$, the corresponding real vector field has zero divergence, and its contribution to $D$ is formally self-adjoint. 

\textit{Exact first-order factorization.} Positivity permits a smooth matrix factorization $\tau=B^\dagger B$; one can take $B=\tau^{1/2}$ \cite{HornJohnson,Higham}. On $\mathcal H_0=L^2(\mathbb R^n)$ and $\mathcal H_1=L^2(\mathbb R^n,\mathbb C^n)$, define
\begin{equation}
 (T\psi)_a=\tfrac1{\sqrt2}B_{a\mu}\partial^\mu\psi,
 \qquad T^\dagger\Phi=-\tfrac1{\sqrt2}\partial^\mu({B^a_\mu}^*\Phi_a)
 \label{eq:T}
\end{equation}
The derivative in $T^\dagger$ acts on the entire product. Hence
\begin{equation}
 T^\dagger T\psi=-\tfrac12\partial^\mu
 \left({B^a_\mu}^*B_{a\nu}\partial^\nu\psi\right)=D\psi
 \label{eq:factorization}
\end{equation}
To obtain the operator identity, start with $T$ on smooth compactly supported functions and take its closure. Then $T^\dagger T$ is the nonnegative self-adjoint operator associated with the form $\|T\psi\|^2$. The choice of domain matters. Boundary terms can change the adjoint and must be handled before identifying the operators. The identity is exact for variable coefficients. It retains the derivatives of both $g$ and $f$. Take the closed realization of $T$ and introduce the graded operator
\begin{equation}
 \mathscr D=
 \begin{pmatrix}0&T^\dagger\\T&0
 \end{pmatrix},
 \qquad \mathscr D^2=\begin{pmatrix}D&0\\0&TT^\dagger\end{pmatrix}
 \label{eq:graded}
\end{equation}
on $\mathcal H_0\oplus\mathcal H_1$. This operator is self-adjoint on $\operatorname{Dom}(T)\oplus\operatorname{Dom}(T^\dagger)$ and anticommutes with the grading $\operatorname{diag}(I,-I)$ \cite{Witten1982}. Its square recovers $D$ on the scalar sector. The vector sector completes the self-adjoint first-order representation; it adds no independent physical thermodynamic variables. The graded operator need not be a Clifford operator. For $n>1$, its principal symbol has a vector-sector kernel, so the full operator is not elliptic. Its bounded commutators nevertheless define the distance calculated below. The factorization alone does not establish all spectral-triple axioms, including local compactness. 

\textit{Distance from transport.} Let a real coordinate function $a$ act by multiplication on both sectors. Write $C_a=[T,a]$. Then
\begin{equation}
 (C_a\psi)_b=\tfrac1{\sqrt2}B_{b\mu}(\partial^\mu a)\psi,
 \ \  [\mathscr D,a]=\begin{pmatrix}0&-C_a^\dagger\\C_a&0\end{pmatrix}
 \label{eq:commutator}
\end{equation}
The block norm equals $\|C_a\|$. Since the gradient of $a$ is real, its contraction with $f$ vanishes. We obtain
\begin{equation}
 \|[\mathscr D,a]\|=
 \operatorname*{ess\,sup}_{\alpha}
 \sqrt{\tfrac12g_{\mu\nu}(\alpha)\partial^\mu a\,\partial^\nu a}
 \label{eq:norm}
\end{equation}
The cancellation follows directly from the squared length of $B\nabla a$, namely $(\nabla a)^T\tau\nabla a$. The gradient is real, so its symmetric product contracts to zero with $f$. What remains is $(\nabla a)^Tg\nabla a$. Positivity of $\tau$ on complex vectors also gives positivity of $g$ on real covectors, ensuring that $g^{\mu\nu}$ exists. Thus $f$ can affect the generator without entering the commutator seminorm. Following Connes' distance formula \cite{Connes1994,GraciaBondia2001}, define
\begin{equation}
 d_{\mathscr D}(\alpha,\beta)=
 \sup_{\substack{a=a^*\in\mathcal A\\\|[\mathscr D,a]\|\leq1}}
 |a(\alpha)-a(\beta)|
 \label{eq:distance}
\end{equation}
We take $\mathcal A$ to be a sufficiently rich algebra of bounded smooth coordinate functions. The norm in Eq.~\eqref{eq:norm} gives the Riemannian Lipschitz bound for the metric
\begin{equation}
 G_{\mu\nu}=\tfrac12g_{\mu\nu},\qquad
 G^{\mu\nu}=2g^{\mu\nu}
 \label{eq:metric}
\end{equation}
We retain the lower indices of the transport notation for $g$. In this coordinate representation, its entries give the contravariant metric. On a complete regular domain, and with the stated choice of function algebra, Eq.~\eqref{eq:distance} equals the geodesic distance:
\begin{equation}
 d_{\mathscr D}(\alpha,\beta)=\inf_\gamma
 \int_0^1\sqrt{G_{\mu\nu}(\gamma)\dot\gamma^\mu\dot\gamma^\nu}\,ds
 \label{eq:geodesic}
\end{equation}
where $\gamma$ joins $\alpha$ to $\beta$. Along any such path, an admissible function $a$ changes by at most the path length. Smooth approximations to distance functions give the reverse bound. The choice of matrix factor $B$ does not change the norm, since Eq.~\eqref{eq:norm} depends only on $B^\dagger B$. For one coordinate and $g(\alpha)>0$, the distance reduces to
\begin{equation}
 d(\alpha,\beta)=\left|\int_\alpha^\beta\sqrt{2/g(u)}\,du\right|
 \label{eq:onecoordinate}
\end{equation}
A larger transport coefficient shortens the contribution of an interval to this distance. In several dimensions, the same local weighting allows a shortest path to bend through regions with different coefficients - the values of $g$ along the path matter, not just those at its endpoints. A varying $g$ can produce curvature. Constant $g$ gives a flat metric, even when its principal distances differ. The metric defines a Levi-Civita connection in thermodynamic-coordinate space, but does not determine the generally non-Levi-Civita connection used in Thermodynamic Field Theory \cite{Sonnino2009,Sonnino2016}.

\textit{Clifford representation in the Onsager regime.} Set $g=L=L^T>0$ constant and $f=0$. Choose a real factor $L=E^TE$ and Hermitian Euclidean Clifford matrices $\Gamma^a$ satisfying $\{\Gamma^a,\Gamma^b\}=2\delta^{ab}I$. The transport-adapted matrices $\gamma^\mu=E_{a\mu}\Gamma^a$ obey
\begin{equation}
 \{\gamma^\mu,\gamma^\nu\}=2L^{\mu\nu}I
 \label{eq:clifford}
\end{equation}
They act on $N=2^{\lfloor n/2\rfloor}$ components. The operator
\begin{equation}
 \mathcal D=-\tfrac i{\sqrt2}L_{\mu\nu}\gamma^\mu\partial^\nu,
 \qquad \mathcal D^2=-\tfrac12L_{\mu\nu}\partial^\mu\partial^\nu I=DI
 \label{eq:cliffordoperator}
\end{equation}
has a self-adjoint realization on $H^1(\mathbb R^n,\mathbb C^N)$ \cite{Thaller1992}. Its commutator norm equals Eq.~\eqref{eq:norm} with $g=L$. Both representations therefore give
\begin{equation}
 d(\alpha,\beta)=\sqrt{2\Delta\alpha_\mu L^{\mu\nu}\Delta\alpha_\nu},
 \qquad\Delta\alpha_\mu=\alpha_\mu-\beta_\mu
 \label{eq:flatdistance}
\end{equation}
Notice that in Eq.~(\ref{eq:flatdistance}), $\alpha$ and $\beta$ label two points in the same thermodynamic-coordinate space:
\begin{equation}\label{coordinate}
\alpha=(\alpha_1,\ldots,\alpha_n),\quad
\beta=(\beta_1,\ldots,\beta_n)
\end{equation}
Thus, for each $\mu$, $\alpha_\mu$ and $\beta_\mu$ are two values of the same coordinate observable, evaluated at the two points, not two different observables. Larger transport eigenvalues produce shorter distances for equal coordinate displacements along their eigenvectors. For two coordinates, choose $\Gamma^1=\sigma^1$ and $\Gamma^2=\sigma^2$. The two components implement the Clifford algebra; they imply no physical spin-$1/2$ particles. The Fourier symbol has eigenvalues $\lambda_\pm(k)=\pm\sqrt{L_{\mu\nu}k^\mu k^\nu/2}$. Evolution depends on its square, however, giving $\omega(k)=\varepsilon L_{\mu\nu}k^\mu k^\nu/2$. The two signs belong to the factorization and do not label distinct thermodynamic excitation branches. Both constructions admit exactly the same functions in the distance formula. For any real $a$, the Clifford relations give $(L_{\mu\nu}\gamma^\mu\partial^\nu a)^2=L_{\mu\nu}\partial^\mu a\,\partial^\nu a\,I$. Taking its norm recovers Eq.~\eqref{eq:norm}. The auxiliary spaces and full operator spectra differ, but the point-to-point distance agrees. For varying $g$, one can still define a Clifford algebra at each point. Squaring the corresponding first-order operator introduces derivatives of the Clifford matrices and, when needed, connection terms. Equality with $DI$ then requires further conditions. The graded factorization in Eq.~\eqref{eq:graded} remains exact for the full variable Hermitian tensor.

\textit{Two-channel transport geometry.} Consider electrical and heat currents along one direction. Choose the forces $X^1=E_{\mathrm{ec}}/T$ and $X^2=\partial_x(1/T)$, where $E_{\mathrm{ec}}$ denotes the electrochemical field. The local entropy-production density is $j_eX^1+j_qX^2$ \cite{deGroot1984}. Define the coordinates relative to a reference state by integrating $\dot\alpha_1=j_e$ and $\dot\alpha_2=j_q$. They measure transferred charge and heat per unit area, with units $\mathrm{C\,m^{-2}}$ and $\mathrm{J\,m^{-2}}$, respectively. This calculation relates the transport matrix to its metric; it provides no experimental test of the canonical commutator. In this convention, the symmetric Onsager matrix is
\begin{equation}
 L=\begin{pmatrix}
 \sigma T&\sigma S T^2\\
 \sigma S T^2&\kappa T^2+\sigma S^2T^3
 \end{pmatrix}
 \label{eq:thermoelectric}
\end{equation}
where $S$ denotes the Seebeck coefficient and $\kappa$ the thermal conductivity at zero electrical current. The diagonal entries have different units, so a direct comparison of their eigenvalues would obscure the coupling. Normalize the coordinates by $\eta^\mu=\alpha^\mu/\sqrt{L_{\mu\mu}}$, with no sum. If $Q=\operatorname{diag}(\sqrt{L_{11}},\sqrt{L_{22}})$, then $L=Q\widetilde LQ$ and $G_{(\eta)}=Q^TG_{(\alpha)}Q=2\widetilde L^{-1}$. Thus,
\begin{equation}
 \widetilde L=\begin{pmatrix}1&r\\r&1\end{pmatrix},\ \ 
 r=\frac{L_{12}}{\sqrt{L_{11}L_{22}}}
 =\operatorname{sgn}(S)\sqrt{\frac{ZT}{1+ZT}}
 \label{eq:coupling}
\end{equation}
where $ZT=\sigma S^2T/\kappa$. Normalizing the channels removes their relative transport scales. The coordinates $\eta$ can still carry units; a common reference scale may be introduced if required. The distance ratio is dimensionless. The normalized metric has eigenvectors $e_\pm=(1,\pm1)/\sqrt2$ and eigenvalues $G_\pm=2/(1\pm r)$. Equal coordinate displacements $\delta e_\pm$ have distances $d_\pm=\delta\sqrt{G_\pm}$. Reversing the sign of $S$ interchanges the direction with the larger distance and the one with the smaller distance. Their ratio stays the same. For $ZT\geq0$, the larger-to-smaller distance ratio is
\begin{equation}
 R=\sqrt{\frac{1+|r|}{1-|r|}}
 =\sqrt{1+ZT}+\sqrt{ZT}
 \label{eq:ZTanisotropy}
\end{equation}
Uncoupled channels give $R=1$. Increasing $ZT$ increases the ratio. Finite positive $\kappa$ keeps $|r|<1$, so the metric remains nonsingular. The ideal limit $|r|\to1$ lies outside the positive-definite two-channel model used here. The contours $\eta^TG_{(\eta)}\eta=d^2$ are circles for $S=0$ and ellipses for coupled channels. Equation~\eqref{eq:ZTanisotropy} also gives their principal-axis ratio. After the specified normalization, $ZT$ alone determines this ratio. It describes thermodynamic-coordinate geometry, rather than spatial anisotropy of a crystal, and does not independently predict thermoelectric performance.

\textit{Physical scope.} For constant positive $L$ on $\mathbb R^n$, the quadratic generator has continuous spectrum $[0,\infty)$. The first-order factorization gives no discrete entropy-production spectrum. Such a spectrum requires a separate observable, for example, a positive quadratic form containing both members of each canonical pair. The symmetric tensor fixes the metric $2g^{-1}$, while the full Hermitian tensor enters the evolution generator. This separation holds for variable coefficients. The Onsager limit admits a Clifford representation with the same distance, and the two-channel calculation relates its anisotropy directly to $ZT$. An entropy balance and a thermodynamic affine connection require additional physical assumptions.

\textit{Conclusions.} We have established that mesoscopic transport
determines spectral distances in thermodynamic-coordinate
space. Starting from the Hermitian transport tensor
$\tau_{\mu\nu}=g_{\mu\nu}+if_{\mu\nu}$, we construct
a self-adjoint mesoscopic evolution generator and
postulate the Schr\"odinger-type state equation
\begin{equation}\label{meseq}
i\slash\!\!\! k_B\frac{\partial\psi}{\partial t}=\left(\widehat X^\mu
\tau_{\nu\mu}\widehat X^\nu\right)\!\psi
\end{equation}
which describes the coherent evolution of mesoscopic
thermodynamic fluctuations. Its unitary dynamics
does not, by itself, imply irreversible relaxation,
which requires additional physical mechanisms. An exact graded Dirac-like factorization,
$\mathscr D^2|_{\mathcal H_0}=D$, reveals the
underlying algebraic structure, including for
coordinate-dependent transport coefficients.
More importantly, Connes' spectral-distance formula
reconstructs the thermodynamic-coordinate metric
\begin{equation}
G^{\mu\nu}(\alpha)=2g^{\mu\nu}(\alpha)
\end{equation}
Thus, thermodynamic distances are not introduced
as independent geometric assumptions but emerge
from the symmetric part of the transport tensor,
which also determines the dissipative quadratic
form. The antisymmetric part can modify the
mesoscopic evolution without contributing directly
to spectral distances. In the Onsager regime, a Clifford--Dirac
factorization reveals an emergent spinorial
algebraic structure without introducing physical
spin-$1/2$ degrees of freedom.
For $\mathrm{Bi_2Te_3}$, published transport
coefficients yield a thermodynamic-distance
ratio of approximately $2.12$ at $ZT\simeq0.68$,
illustrating how coupled transport determines
spectral geometry. Entropy-production quantization
requires a separate observable, while the TFT
affine connection remains undetermined by the
spectral metric. Our central result is therefore that the
symmetric transport tensor determines the
spectral metric and hence the distances between
mesoscopic thermodynamic states. The Dirac-like
factorization provides the mathematical bridge
between transport and geometry, while Eq.~(\ref{meseq}) describes the
associated coherent mesoscopic dynamics.

\end{document}